# THE PHYSICS LABORATORY WORKS – INDIVIDUALIZED COMPUTER SIMULATIONS

A.D.Zaikin, I.I.Suhanov

Novosibirsk State Technical University, Novosibirsk, Russia

The physics laboratory-works creating and operating computer simulations experience is described.

A significant amount of laboratory works can be classified as a "black box". The studied physical phenomenon is hidden from direct observation, the control is carried out by means of electrical measuring devices. It is difficult to distinguish physical reality from its imitation when performing such work, so the virtualization of this one does not require realistic images. The schematic representation of the laboratory installation greatly simplifies the process of creating a simulator.

A unique set of installation parameters is formed for each student performing laboratory work on the simulator. These parameters are stored in Google Tables. Their transfer to the laboratory works html-template is carried out in encrypted form through the Google Apps script service.

The simulator-parameters individualization contributes to the independence of the student's work when performing laboratory measurements in the conditions of the distance learning.

## 1. Introduction

The development of distance education, having a rich history, has become avalanche-like in the last two years. The teaching community suddenly faced issues of a methodological, organizational and scientific nature that were previously unknown to most.

The specifics of higher school disciplines are such that the transition was almost painless for some of them, but for others it required a radical breakdown of the educational process, and some ones are simply impossible in a distance format.

The physics course for engineers involves various forms of interaction with the student. Apparently, it is most difficult to translate a laboratory workshop into a remote format, although virtual and remote laboratory work has been practiced before. Such developments were based on the enthusiasm and the availability of appropriate skills of the teacher. The pandemic has made this direction mandatory.

Laboratory practical training in physics is very important and instructive for the full-fledged understanding formation of the physical phenomena essence in a student. The laboratory work performed together with the study of theoretical material helps to feel the connection of a physical experiment with theoretical concepts, to understand and assimilate new knowledge more deeply. In addition, the student gets the skill of working with devices, learns to measure and process measurement results, build dependencies and graphs.

Here is a brief overview of existing solutions in this area.

Laboratory work performed at home [1] is one of the ways to conduct remote classes. In such works, the student is supposed to use not specialized equipment, but devices that have a household purpose, for example, scales, a thermometer, 3D glasses, a laser pointer, a smartphone, etc. Methodological recommendations for home laboratory work usually regulate only the main stages of their implementation. The student is given more freedom of action than working in a training laboratory. The studied physical phenomenon is real and tangible.

The project of an Internet laboratory with remote control for conducting experiments in physics is very interesting [2]. The user, through a browser, gets access to a webcam and electronic components located in a remote laboratory, such as a USB K8055 board, an Arduino with an Ethernet Shield, a Raspberry Pi, to control a real physical device. The natural limitation of the project is the throughput capacity. Developers solve this problem, which occurs when working on any real laboratory installation, by booking access time. If the installation is free, then you can immediately start measuring.

Computer simulations imitate a physical phenomenon allowing you to visualize processes that are inaccessible to a direct observation, scale them in space and time. In classes, both lectures and seminars, simulators are used as animated illustrations. This form of presentation of the material is more effective than static drawings and can often compete with full-scale demonstrations.

Usually a computer simulation is a complete program created on the basis of a certain technology, it can be a Java applet or an Adobe Flash file. It should be noted that these technologies are outdated and are no longer supported by manufacturers. A demonstration of simulators implemented using these technologies on the user's device is a often difficult and sometimes unrealizable task. The HTML5 markup language is considered a modern tool, which allows you to run a simulation in modern browsers on any device.

The PhET project [3], implemented by the University of Colorado, is one of the most well – known and well-developed in this field and offers a wide range of interactive scientific and mathematical simulations.

A simulation model of a physical process is implemented in such a way that it often includes a more or less realistic image of measuring devices and other equipment. Such simulations can be adapted as a stand for laboratory work, exerting a certain amount of ingenuity [4].

A separate class of solutions is a complete program module, executable in a specific operating system [5]. Almost always, this operating system is the Windows OS family. The authors are not aware of cross-platform solutions for such products. The student installs the program after downloading it on the computer. The installation process is not always smooth. The reasons for the difficulties are very diverse: outdated and unsupported versions of the operating system, the presence of the required archiver and application libraries, the interaction of the program with antivirus protection, the lack of rights to install the software product.

As the most modern browsers adhere to international standards in the field of data processing and display, performing a laboratory work simulation directly in the browser allows you to use any device: a computer, tablet, smartphone. This approach seems to be optimal.

The Internet resource [6] contains more than two hundred virtual laboratory works and demonstrations in all sections of physics. However, for them to work, the Adobe Flash Player plugin, not currently supported, must be installed in the user's browser. These circumstances significantly limit the possibilities of using this product. The complexity of the launch may exceed the complexity of performing the laboratory work itself.

**2. Proposed solutions**

The experience of transferring the course of general physics of engineering specialties of the technical University to the remote format of a laboratory workshop follows below.

It is difficult to overestimate the visibility of some very simple physical experiments from the point of view of implementation. The phenomena arising from the free oscillations of coupled pendulums, the standing wave formation on a string, the laser light interference and diffraction, the light beam quenching by two transparent plates of polarizers border on magic. With the virtualization of such physical phenomena the loss of direct participation in the process is inevitable, despite all the power of modern technologies.

However within the framework of a standard laboratory practice in physics there is a significant number of works that can be classified as a "black box". As a rule these are ones related to electrical measurements.

An example of such work is the electron-specific-charge determination by the magnetron method, where the processes are monitored by means of measuring devices: a voltmeter and ammeters. It is extremely difficult in fact to distinguish a physical reality from its imitation by registering the readings of instruments when performing this work.

In our opinion, the virtualization of such works does not require a naturalistic image of the devices, a conditional schematicity is quite enough, thus significantly simplifies the simulator development technology. In order to familiarize the student with real devices, if other options are unavailable, accompanying video clips will be suitable.

The independent performance of laboratory work by a student is a problem that has become relevant in the distance learning conditions. The parameters of the virtual laboratory installation controlled by the teacher and individual for each student are one of the possible answers to this challenge.

As the first step, five laboratory works that are in demand in the current educational process were selected from an extensive laboratory workshop for virtualization. Each of them corresponds to the definition of a "black box". A detailed description of the work is given below.

The creation of all simulators was carried out using a single technology developed for this task. It is based on browser solutions using the HTML5 markup language, which simplifies the creation and management of multimedia objects without the need to use third-party plugins.

On the corresponding web page of the university portal [7] the student forms a request to perform a specific laboratory work. To do this, he fills in the fields determining the number of the study group, last name and password (Figure 1). The student who is allowed to perform the work receives the password from the teacher.

**Laboratory work No. 7**
**Studying the Boltzmann distribution**
This work in the laboratory practice of the department

Group: EN2-01 ▾  Surname: Ivanov    Password: ****    Execute

Fig. 1

When the "Execute" button is clicked, a PHP script is launched that loads an HTML template of the laboratory work containing diagrams, drawings, control elements and PHP scripts that implement a mathematical model of the physical process under study.

The general elements of the template are supplemented with the student's personal data and the installation parameters individualized for him. This data is contained in a Google table that includes group lists of students and a logbook sheet in which the student's work is logged: login date, group, last name, IP address and browser.

The PHP template script gets access to the Google Table through the Google Apps Script, published as a web application. Google Apps Script is a development platform for creating small applications that integrate with Google Workspace.

Firstly the Google Apps script checks whether the entered personal data matches the data contained in the table. A mismatch leads to a denial of access. If this stage is successfully passed, a string of individual parameters is formed and returned in encrypted form to the template of the laboratory work.

Some of the individualized installation parameters are hidden, inaccessible to observation during the execution of the work. They are determined by the student based on the processing measurements results. On the contrary, others are open. These ones can be the value of the capacitance, inductance, electromotive force, the maximum value of the current. These parameters are displayed on the screen, fall into the laboratory work report. They are individual therefore they allow the teacher to identify each student not only by the last name.

The cathedral workshop, which contains a significant number of real laboratory works on various branches of physics, is presented in [8]. The principle of selecting works for virtualization, defined above, was combined with the requirements of the current educational process. Next, we will describe the features of the five developed simulators.

### 3. The direct multiple measurements processing

Laboratory practical training in physics usually begins with the exploration the direct-multiple-measurements processing methods by the student. The ways of implementing a specific measurement procedure are very diverse. Measuring the spring pendulum oscillation period, the isotope source intensity, the ball diameter, the electric circuit parameters is only a small list of possible options. In the workshop [8] for these purposes the collision time of metal balls is measured with a microsecondometer.

We believe the use of computer simulation for carrying out such laboratory work is quite justified.

A sample of 50 numbers distributed according to the normal law is pre-generated for each student. The average value and the standard deviation are individual.

The methods of generation can be very different, for example, it can be carried out directly in the Google table.

We get the required number of values distributed according to the normal law by writing a formula in one of the cells and dragging the autofill marker.

You should not use the resulting sample blindly, without visualization. The sample of 50 values is not very representative, as a result the histogram does not always take canonical forms, losing its visibility. An example of an unsuccessful histogram is shown in Figure 2 on the left. For educational purposes, it is preferable to use the option shown on the right. An acceptable option is obtained by simply recalculating the entered formulas.

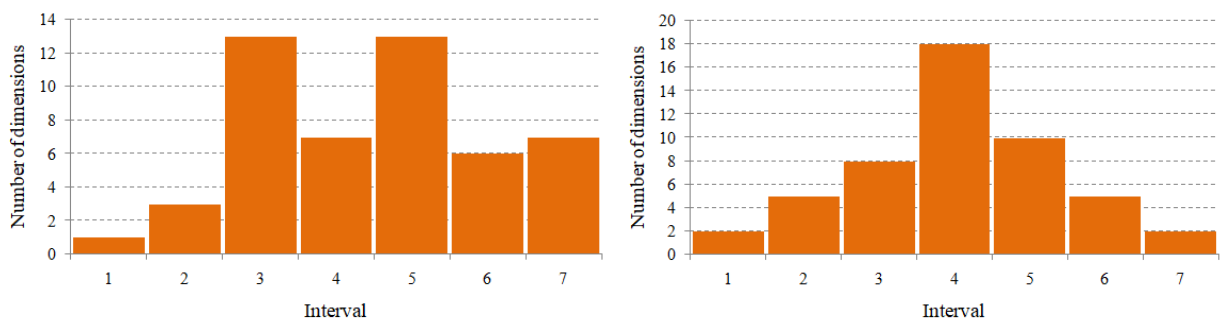

Fig. 2

The template of this work contains a short video fragment of the two balls collision, as well as a control element – the "Start" button, Figure 3. When the button is pressed, the video fragment is played, and the time of the collision is displayed on the stopwatch display.

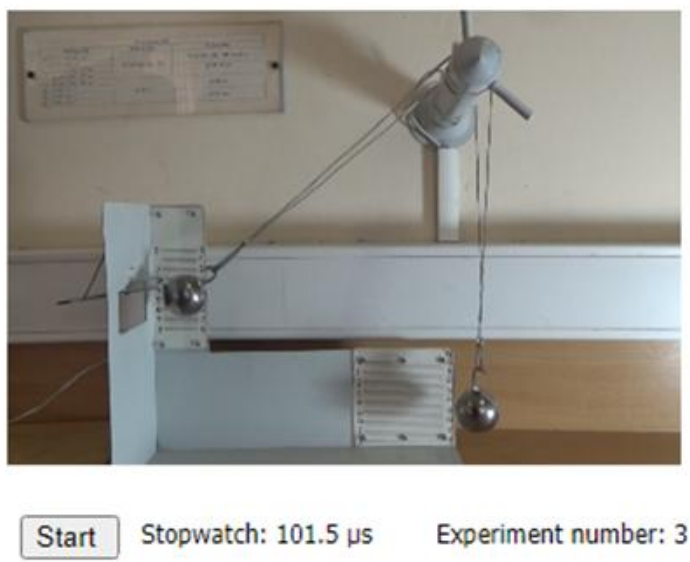

Fig. 3

The student's laboratory work consists in pressing the "Start" button fifty times and registering the microsecond meter readings in the measurement table. The subsequent processing of the measurement results is carried out in accordance with the requirements of the methodological recommendations.

## 4. Study of the Boltzmann distribution

The object of research in this work is an electrons cloud located between the cathode and the anode of an electric vacuum diode. The electrons emitted by the cathode form an ideal gas whose temperature is equal to the cathode's one. The concentration of the electron gas is constant in the absence of an electric field. In the presence of an external force field the concentration becomes spatially inhomogeneous. The reverse inclusion of the diode (a positive potential is applied to the cathode and a negative potential is applied to the anode) reduces the concentration of electrons from the cathode to the anode according to the Boltzmann distribution.

The computer simulation template along with the control elements contains a schematic diagram of the installation, Figure 4. The scheme includes an ammeter, a voltmeter measuring the voltage between the anode and the cathode, and a rheostat.

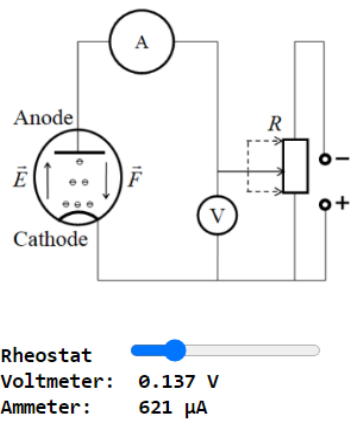

Fig. 4

The slider rheostat can change the readings of the voltmeter $U$ from 0 to 700 mV in increments of 1 mV.

The anode current $I$ in this case changes according to the expression

$$I = I_{max} \exp(-qU/kT)$$

A typical dependence of the anode current versus the voltage on a semi-logarithmic scale is shown in Figure 5. A student should build this dependence based on the results of measurements.

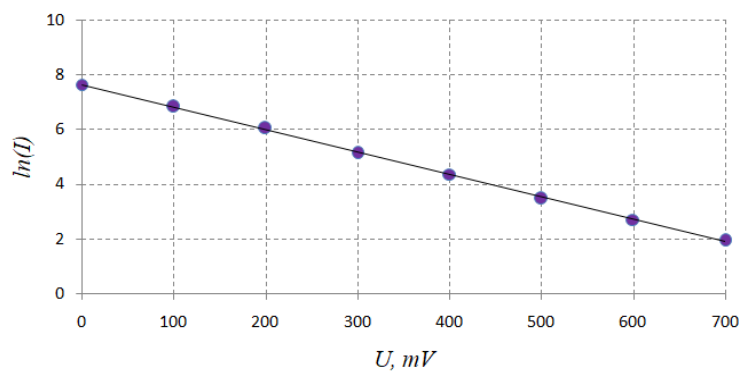

Fig. 5

The values of the current in the absence of an electric field $I_{max}$ and the temperature of the electron gas $T$ are individual in each laboratory work. The range of operating values of the current is 2100-2400 µA, and the temperature is 1100-1400 K.

The personification of the student is carried out by the value $I_{max}$, this is an open parameter, whereas the temperature $T$ is calculated during the execution of the work and cannot be an unambiguous sign.

To make the measurements more natural the calculated ammeter readings are modified. The values of the anode current $I$ are replaced by the values obtained by generating random numbers evenly distributed in the interval (*0.8I; 1.2I*). The only exception is the value $I(0) = I_{max}$ used for personalization remains unchanged.

The student, when performing this laboratory work, moves the slider of the rheostat, registers the readings of the ammeter and voltmeter, then constructs the current-voltage characteristic and calculates the temperature of the electron gas.

**5. Study of the operation of a DC power source**

An unbranched DC electrical circuit contains a current source and a resistor with variable resistance. The measuring instruments, ammeter and voltmeter are connected as shown in the diagram (Figure 6). The dependence of the voltage drop on the resistor on the current $U = E - Ir$ is linear. By measuring the current-voltage characteristic evaluation it is possible to determine the electromotive force (EMF) $E$ of the source, its internal resistance $r$, short-circuit current $I_{sc} = E/r$, useful and full power.

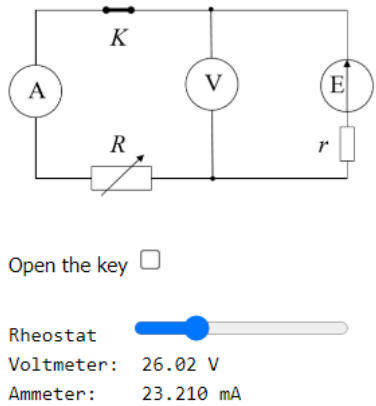

Fig. 6

As part of the simulation it is assumed that the range of EMF is 10-60 V with 0.5 V increments and the range of internal resistance is 400-700 Ohms. Then the short-circuit current range is 15-150 mA.

Moving the rheostat slider, we set a discrete real number $n_i$ in the interval [0; 1] with a step 0.001. Then the values of the current are determined by the expression $I_i = I_{\kappa3} n_i$, and the voltage $U_i = E - rI_i$.

In total, 1000 values of the measured parameters can be obtained in the simulation. If the number $i > 900$, an over current is registered and the ammeter displays a message about it instead of the readings. If the value of i is<100, then the values are taken at the point $i = 100$, corresponding to the residual resistance of the rheostat. At a values of $i > 300$ the voltmeter readings are modified in the range of ±1% of the calculated value by means of a random number generator.

The EMF value (open parameter) and the internal resistance of the source (closed parameter) are personalized and transmitted to the simulation template for a specific student. The identification of the student is carried out according to the readings of the voltmeter with the open key, corresponding to the EMF value.

**6. The electron specific charge determination**

The central element of a laboratory installation for measuring the electron specific charge is a magnetron, an electro-vacuum triode placed in a solenoid. The dependence of the anode current versus the solenoid one allows you to calculate the specific charge. The computer simulation template includes a schematic diagram of the installation and control elements, Figure 7. These include two ammeters, one in the solenoid circuit, another in the anode one, a voltmeter measuring the anode - cathode voltage, and two rheostats.

The anode-cathode voltage is set using the rheostat $R_1$. The rheostat $R_2$ regulates the current in the solenoid circuit from 0 to 2 A in increments of 1 mA. The anode current is measured by an ammeter $A_1$, and the current in the solenoid circuit is measured by an ammeter $A_2$.

The analysis of the motion of electrons in a magnetron leads to the following conclusion. The ideal anode - solenoid currents dependence $I_a(I_s)$ has the form of a step. The solenoid current which stops the anode one, is called a cutoff. In the conditions of a real experiment the step is eroded primarily due to the magnetic field inhomogeneity along the axis of the finite length solenoid. On real curves the cutoff current $I_C$ is determined from the condition $I_a(I_{omc}) = I_0/2$.

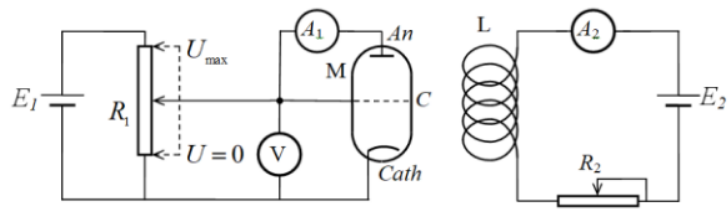

**Control elements**

Rheostat $R_1$         Voltmeter V:  62 V
Rheostat $R_2$
Ammeter $A_1$:  13.6 mA
Ammeter $A_2$:  0.732 A
Magnetron parameters: l=(62±2) mm, R= (2.8±0.2) mm, N=1900

Fig.7

We will construct a correspondence between the anode-cathode voltage $U$ and the cutoff current $I_C$, as well as the anode current at zero current in the solenoid circuit $I_0$.

To calculate the anode current, we use the three-halves-power law

$$I_0 = aU^{3/2},$$

where $a$ is the perveance, a coefficient that depends on the configuration, dimensions and material of the electrodes.

The simulation parameters are selected so that the quantitative results of the experiments correspond to the operating modes of the real installation. If an interval $0.05 \leq a \leq 0.0995$ is choosed for the perveance, then at voltages (60-80) V the values of the anode current will be in the range (23-71) mA.

Let the number of turns of the solenoid be $N$, the length is $l$, and the radius is $R$, then these parameters and the electron specific charge are related by the ratio

$$\frac{e}{m} = \frac{8l^2}{(K\mu_0 NR)^2} \frac{U}{I_C^2}$$

The correction factor $K$ is introduced to account for the inhomogeneity of the magnetic field.

For the cut-off current of the ideal step we obtain $I_{omc} = b\sqrt{U}$, where $b = \frac{2l}{K\mu_0 NR}\sqrt{\frac{2m}{e}}$.

Assuming the parameters of the magnetron as follows: $R = (2.8 \pm 0.2)\,mm$, $l = (62 \pm 3)\,мм$, $K = 0.65$, and taking into account that $\mu_0 = 4\pi \cdot 10^{-7}\,H/m$ and $e/m = 1.75882 \cdot 10^{11}\,C/kg$, we get $b = 182.82885/N$.

The constants $a$ and $N$ are individual and transmitted to the template for a specific student. The number of turns of the solenoid is limited by the interval $1900 \leq N \leq 4000$.

The known constants $a$ and $N$ and the exposed cathode-anode voltage $U$ uniquely determine the values of $I_0$ and $I_C$ in the constructed model.

To calculate the value of the anode current, the function was selected

$$I_a(I_s) = I_0 \frac{\exp\left(-\frac{I_C}{D}\right) + 1}{\exp\left(\frac{I_s - I_C}{D}\right) + 1},$$

which blurs the perfect step.

Here $D = cI_C$ is the width of the blur area, it was assumed that $c = 0.15$. It is true $I_a(0) = I_0$ for the constructed function, its graph is shown in Figure 8.

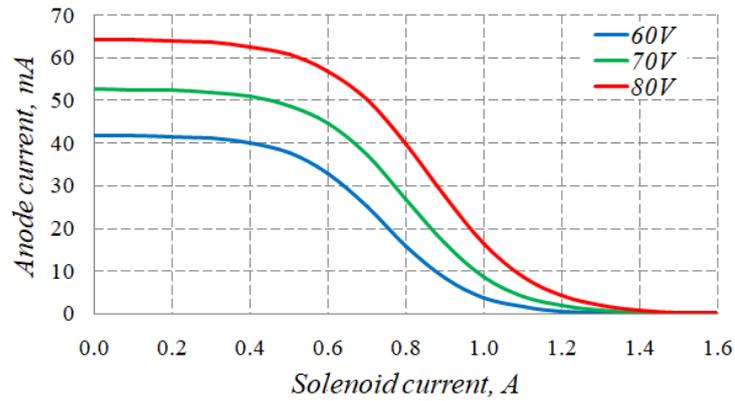

Fig. 8

To make the ammeter $A_1$ readings realistic in the range (10-90)% of the maximum value was made noisy by the random number sensor.

The open parameter in this work is the number of the magnetron turns $N$, and the closed parameter is the perveance $a$.

Performing the work, the student sets the anode voltage with the rheostat $R_1$, moves the rheostat $R_2$ slider, registers readings of ammeters $A_1$, $A_2$, then builds the anode - solenoid currents dependence.

### 7. Forced oscillations in the oscillatory circuit

The voltage on the capacitor of an oscillatory circuit containing a series-connected harmonic signal generator depends on the signal frequency. The phenomenon of resonance can be studied by measuring the capacitor voltage dependence on frequency. The schematic diagram of the laboratory installation is shown in Figure 9.

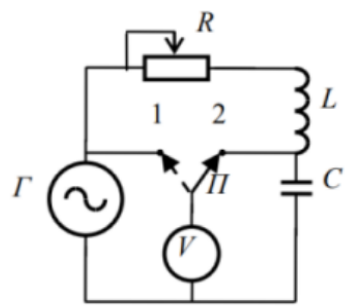

**Control elements**

Generator ▬▬▬●▬▬ Frequency: 34520 Hz
Rheostat ▬▬▬▬▬●▬ Resistance: 294 Ω
Voltmeter: 0.076 V
Circuit parameters: C = (20 ± 1) nF, L = (15.0 ± 0.2) mH, E = 1.000 V

Fig. 9

In addition to the mentioned generator and voltmeter, measuring the voltage, the installation contains a rheostat.

The control elements of the simulation are a slider that changes the frequency of the generator output signal, and a slider rheostat that changes the resistance of the oscillatory circuit.

The following constants are individual and transmitted to the template for a particular student: the electrical capacity of the capacitor $C$, the inductance of the coil $L$ and three resistance values for which resonant curves are taken. The output voltage of the generator $E$ is assumed to be constant and equal to one volt.

In order for the simulation parameters generally correspond to the parameters of the real installation, it is assumed that the range of resistances used is 30-400 Ohms, the capacitance of the capacitor is 20-80 nF, the inductance of the coil is 2-50 mH. Then the resonant frequency of the circuit is within the range of 2920-25200 Hz.

The frequency of the harmonic signal of the generator can be changed in the range of 10-100000 Hz in increments of 10 Hz. The resistance of the slider rheostat in the oscillating circuit changes discretely in increments of 1 Ohm in the range of 10-500 Ohms.

Having set the specified resistance value, the student, by changing the frequency of the generator, measures the amplitude-frequency response $U(\nu)$. The recorded readings of the voltmeter $U$ are calculated according to the known formulas

$$U = \frac{\nu_0^2 E}{\sqrt{(\nu^2 - \nu_0^2)^2 + \beta^2 \nu^2 / \pi^2}} \quad , \quad \beta = \frac{R}{2L} \quad , \quad \nu_0 = \frac{1}{2\pi\sqrt{LC}} \quad .$$

Having constructed a resonance curve based on the measurement results, the student determines the resonance frequency and Q-factor.

### 8. Conclusion

The approbation of the described works took place in the process of studying physics by students of the technical university. The forced transition to distance learning in the pandemic context led to the search for new forms of teacher-student interaction. The students performed laboratory work using any device available to them. It could be both a tablet and a smartphone. The individualization of the parameters of the virtual installation had a beneficial effect on the independence of work. Rare attempts to use other people's results had become immediately obvious.

The experience gained in the process of work will be useful when expanding the list of laboratory work simulators.

# LITERATURE

1. Laboratory work at home //Department of General Physics /MIPT [Electronic resource]. – URL https://mipt.ru/education/chair/physics/news/laboratornye_raboty_v_domashnikh_usloviyakh (accessed: 07.06.021).
2. Internet laboratory with remote control [Electronic resource]. - URL: http://remote-lab.fyzika.net/experiment/02/experiment-2.php?lng=en (accessed: 07.06.021).
3. Interactive modeling for natural sciences and mathematics [Electronic resource]. – URL: https://phet.colorado.edu/ (accessed: 07.06.021).
4. Interactive modeling of physical phenomena in the Phet environment: a method. instructions and collection of tasks / Novosibirsk State Technical University. univ.; comp.: M. P. Sarina, A.V. Topovsky. – Novosibirsk : NSTU, 2014. – 72 p.
5. Virtual laboratory work // Department of General Physics of REF /NSTU [Electronic resource]. - URL: https://ciu.nstu.ru/kaf/of/virutalne_laboratorne_rabot (accessed: 07.06.021).
6. Virtual laboratory work in physics // Moscow State Educational Complex (MGOK) [Electronic resource]. – URL: http://mediadidaktika.ru (accessed: 07.06.021).
7. Computer simulations of laboratory work // Department of Applied and Theoretical Physics / NSTU [Electronic resource]. – URL: http://pitf.ftf.nstu.ru/files/zaikin/LabComSim.html (accessed: 07.06.021).
8. Laboratory workshop on Physics // Department of Applied and Theoretical Physics /NSTU [Electronic resource]. – URL: http://pitf.ftf.nstu.ru/resources/labs/ (accessed: 07.06.021).

# ИНДИВИДУАЛИЗИРОВАННЫЕ КОМПЬЮТЕРНЫЕ СИМУЛЯЦИИ ЛАБОРАТОРНЫХ РАБОТ ПО ФИЗИКЕ

А.Д. Заикин, И.И Суханов

Новосибирский государственный технический университет, г. Новосибирск, Россия

Изложен опыт создания и эксплуатации компьютерных симуляций лабораторных работ по физике. Значительное количество лабораторных работ можно классифицировать как «черный ящик». Исследуемое физическое явление скрыто от непосредственного наблюдения, контроль осуществляется посредством электрических измерительных приборов. При выполнении такой работы отличить физическую реальность от ее имитации затруднительно. Виртуализация подобных лабораторных работ не требует реалистичности изображений. Схематичность представления лабораторной установки существенно упрощает процесс создания симулятора.

Для каждого студента, выполняющего лабораторную работу на симуляторе, формируется уникальный набор параметров установки. Эти параметры хранятся в Google таблицах. Их передача в html-шаблон лабораторной работы осуществляется в зашифрованном виде посредством сервиса скриптов приложений Google Apps.

Индивидуализация параметров симулятора способствует самостоятельности работы студента при выполнении лабораторной в условиях дистанционного обучения.

**Ключевые слова:** лабораторная работа, симулятор, Google таблица, скрипт Google Apps.

## 1. Введение

Развитие дистанционного образования, имеющего богатую историю, в последние два года приобрело взрывной характер. Перед преподавательским сообществом внезапно возникли неведомые ранее большинству вопросы методического, организационного, научного характера.

Специфика дисциплин высшей школы такова, что для некоторых из них переход был почти безболезненным, для других же требовалась коренная ломка учебного процесса, а отдельные учебные курсы просто невозможны в дистанционном формате.

Курс физики для инженерных специальностей предполагает разнообразные формы взаимодействия со студентом. По-видимому, наиболее сложно перевести на дистанционный формат лабораторный практикум. Хотя виртуальные и дистанционные лабораторные работы практиковались и ранее. В основе таких наработок лежал энтузиазм и наличие соответствующих навыков преподавателя. Пандемия сделало это направление обязательным.

Лабораторный практикум по физике весьма важен и поучителен для формирования у студента полноценного представления о сути физических явлений. Выполненная лабораторная работа совместно с изучением теоретического материала помогает ощутить связь физического эксперимента с теоретическими представлениями, глубже понять и усвоить новые знания. В дополнение к этому студент получает навык работы с приборами, учится измерять и обрабатывать результаты измерений, строить зависимости и графики.

Приведем краткий обзор существующих решений в данной области.

Лабораторные работы, выполняемые в домашних условиях [1], – один из способов проведения дистанционных занятий. В таких работах предполагается использование студентом не специализированного оборудования, а устройств, имеющих бытовое предназначение, например весы, термометр, 3D-очки, лазерная указка, смартфон и т.п. Методические рекомендации к домашним лабораторным работам обычно регламентируют лишь основные этапы их выполнения. Студенту предоставляется большая, по сравнению с работой в учебной лаборатории, свобода действий. Изучаемое физическое явление при этом реально и осязаемо.

Весьма интересен проект интернет-лаборатории с дистанционным управлением для проведения экспериментов по физике [2]. Пользователь, посредством браузера, получает доступ к находящимся в удаленной лаборатории вебкамере и электронным компонентам, таким как плата USB K8055, Arduino с Ethernet Shield, Raspberry Pi, для управления реальным физическим прибором. Естественное ограничение проекта – пропускная способность. Эту проблему, возникающую при работе на любой реальной лабораторной установке, разработчики решают путем бронирования времени доступа. Если же установка свободна, то можно сразу приступить к измерениям.

Компьютерные симуляции имитируют физическое явление, позволяя визуализировать недоступные для непосредственного наблюдения процессы, масштабировать их в пространстве и времени. На занятиях, как лекциях, так и семинарах, симуляторы используются в качестве анимированных иллюстраций. Такая форма подачи материала более эффективна, чем статические рисунки и зачастую может конкурировать с натурными демонстрациями.

Обычно компьютерная симуляция – это законченная программа, созданная на базе определенной технологии, это может быть Java-апплет или файл формата Adobe Flash. Следует отметить, что данные технологии являются устаревшими и более не поддерживаются производителями. Демонстрация симуляторов, реализованных с использованием данных технологий, на устройстве пользователя весьма непростая, а иногда и нереализуемая задача. Современным средством считается язык разметки HTML5, позволяющий запускать симуляцию в современных браузерах на любом устройстве.

Один из самых известных и проработанных в этой области – проект PhET [3], реализуемый университетом Колорадо, предлагает широкий набор интерактивных научных и математических симуляций.

Зачастую симуляционная модель физического процесса реализована так, что включает в себя в той или иной степени реалистичное изображение измерительных приборов и иного оборудования. Проявляя известную долю изобретательности [4], такие симуляции можно приспособить в качестве стенда для проведения лабораторной работы.

Отдельный класс решений – законченный модуль, представляющий собой исполняемую в конкретной операционной системе программу [5]. Практически всегда этой операционной системой служит семейство OS Windows. Кроссплатформенные решения для таких продуктов авторам неизвестны. Скачав программу, студент устанавливает ее на свой компьютер и выполняет виртуальную лабораторную работу. Далеко не всегда процесс установки проходит гладко. Причины возникающих трудностей весьма разнообразны: устаревшие и не поддерживаемые версии операционной системы, наличие требуемого архиватора и прикладных библиотек, взаимодействие программы с антивирусной защитой, отсутствие прав на установку программного продукта.

Исполнение симуляции лабораторной работы непосредственно в браузере, а большинство современных браузеров придерживается международных стандартов в области обработки и отображения данных, позволяет использовать любое устройство: компьютер, планшет, смартфон. Такой подход представляется оптимальным.

Интернет-ресурс [6] содержит более двухсот виртуальных лабораторных работ и демонстраций по всем разделам физики. Однако для их работы в браузере пользователя должен быть установлен неподдерживаемый в настоящее время плагин Adobe Flash Player. Данные обстоятельства существенно ограничивают возможности использования этого продукта. Сложности запуска может превышать сложность выполнения самой лабораторной работы.

## 2. Предлагаемые решения

Далее излагается опыт перевода на дистанционный формат лабораторного практикума курса общей физики инженерных специальностей технического университета.

Наглядность некоторых, совсем простых с точки зрения реализации, физических опытов трудно переоценить. Явления, возникающие при свободных колебаниях связанных маятников, образовании стоячей волны на струне, интерференция и дифракция лазерного света, гашении светового пучка двумя прозрачными пластинками поляризаторов граничат с магией. При виртуализации таких физических явлений, несмотря на всю мощь современных технологий, неизбежны потери непосредственного участия в процессе.

Вместе с тем в рамках стандартного лабораторного практикума по физике существует значительное количество работ, которые можно классифицировать как «черный ящик». Как правило, это работы, связанные с электрическими измерениями.

Пример такой работы – определение удельного заряда электрона методом магнетрона. Электровакуумный триод размещен в соленоиде, контроль процессов осуществляется посредством измерительных приборов: вольтметра и амперметра. Фактически, регистрируя показания приборов при выполнении этой работы, отличить физическую реальность от её имитации крайне затруднительно.

На наш взгляд виртуализация подобных работ не требует натуралистичности изображения приборов, вполне достаточно условной схематичности, что существенно упрощает технологию разработки симулятора. А в целях ознакомления студента с реальными приборами при недоступности иных вариантов подойдут сопутствующие видео ролики.

Самостоятельность при выполнении лабораторной работы студентом – проблема, актуализировавшаяся в условиях дистанционного обучения. Контролируемые преподавателем и индивидуальные для каждого студента параметры виртуальной лабораторной установки – один из возможных ответов на данный вызов.

В качестве первого шага из обширного лабораторного практикума для виртуализации было выбрано пять лабораторных работ, востребованных в текущем учебном процессе. Каждая из них соответствует определению «черный ящик». Подробное описание работ приводится далее.

Создание всех симуляторов осуществлялась по единой, разработанной для данной задачи технологии. В основу положены браузерные решения, использующие язык разметки HTML5, упрощающий создание и управление мультимедийными объектами без необходимости использования сторонних плагинов.

На соответствующей web-странице университетского портала [7] студент формирует запрос на выполнение конкретной лабораторной работы. Для этого он заполняет поля (Рисунок 1), определяя номер учебной группы, фамилию и пароль. Пароль студент, допущенный к выполнению работы, получает у преподавателя.

**Лабораторная работа №7**
**Изучение распределения Больцмана**

Данная работа в лабораторном практикуме кафедры

Группа: ЭН2-01 Фамилия: Иванов Пароль: ****** Выполнить

Рисунок 1

При нажатии кнопки «Выполнить» запускается PHP-скрипт, который загружает html-шаблон лабораторной работы, содержащий схемы, рисунки, управляющие элементы и PHP-скрипты, реализующие математическую модель исследуемого физического процесса.

Общие элементы шаблона дополняются персональными данными студента и индивидуализированными для него параметрами установки. Эти данные содержит Google таблица, включающая в себя групповые списки студентов и лист – журнал входов, в котором логируется работа студента: дата входа, группа, фамилия, IP адрес и браузер.

PHP скрипт шаблона получает доступ к Google таблице посредством скрипта Google Apps, опубликованного как web-приложение. Google Apps Script – это платформа разработки, которая позволяет создавать небольшие приложения интегрированные с Google Workspace.

Первое, что делает скрипт Google Apps – проверяет соответствие введенных персональных данных содержащимся в таблице. Несоответствие приводит к отказу в выполнении работы. Если этот этап успешно пройден, то формируется строка индивидуальных параметров, которая в зашифрованном виде возвращается в шаблон лабораторной работы.

Некоторые из индивидуализированных параметров установки являются скрытыми, недоступными для наблюдения в процессе выполнения работы. Они определяются студентом по результатам обработки измерений. Другие, наоборот, открытыми. Такими параметрами могут быть величина емкости, индуктивности, ЭДС, максимальное значение силы тока. Эти параметры отображается на экране, попадают в протокол лабораторной работы. Они индивидуальны, следовательно, позволяют идентифицировать каждого студента не только по фамилии.

Кафедральный практикум, содержащий значительное количество реальных лабораторных работ по различным разделам физики, представлен в [8]. Принцип отбора работ для виртуализации, определенный выше, сочетался с требованиями текущего учебного процесса. Далее опишем особенности пяти разработанных симуляторов.

### 3. Обработка результатов прямых многократных измерений

Лабораторный практикум по физике обычно начинается с освоения студентом методов обработки результатов прямых многократных измерений. Способы реализации конкретной измерительной процедуры весьма разнообразны. Измерение периода колебаний пружинного маятника, интенсивности изотопного источника, диаметра шарика, параметров электрической цепи – это лишь малый перечень возможных вариантов. В практикуме [8] для этих целей проводится измерение микросекундомером времени соударения металлических шаров.

Использование компьютерной симуляции для проведения такой лабораторной работы на наш взгляд вполне оправдано.

Предварительно генерируется для каждого студента выборка из 50 распределенных по нормальному закону чисел. Индивидуальными являются среднее значение и стандартное отклонение.

Способы генерации могут быть самыми различными, например, ее можно осуществить непосредственно в Google таблице.

Записав в одной из ячеек формулу и перетаскивая маркер автозаполнения, получим необходимое количество значений, распределенных по нормальному закону.

Не следует использовать полученную выборку вслепую, без визуализации. Выборка из 50 значений не очень представительна, как следствие – гистограмма не всегда принимает канонические формы, теряя свою наглядность. Пример неудачной гистограммы приведен на Рисунке 2 слева. В учебных целях предпочтительнее использовать вариант, приведенный справа. Приемлемый вариант получается простым пересчетом введенных формул.

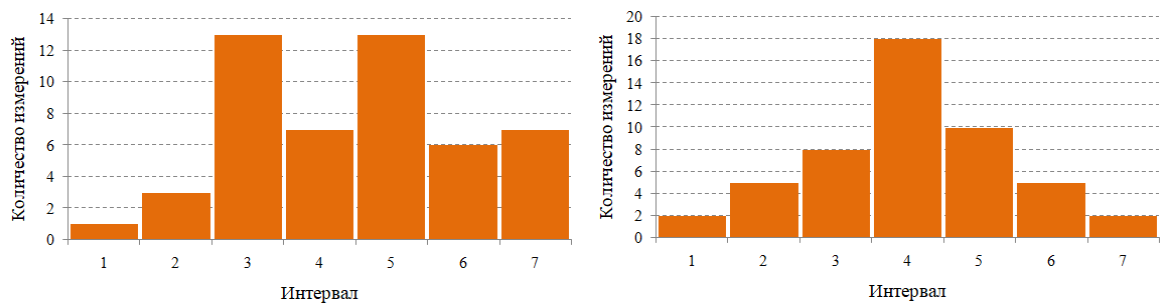

Рисунок 2

Шаблон данной работы содержит короткий видео-фрагмент соударения двух шаров, а также управляющий элемент – кнопку «Пуск», Рисунок 3. При нажатии кнопки проигрывается видео-фрагмент, а на табло секундомера выводится время соударения.

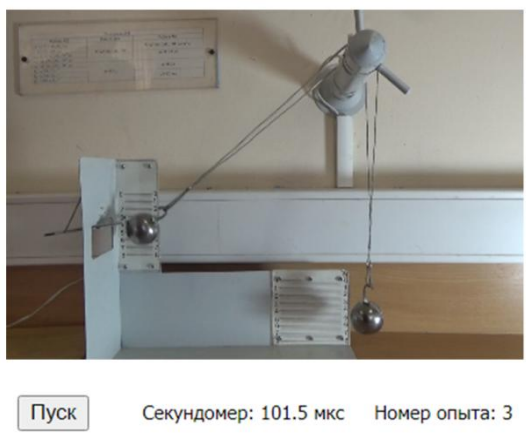

Рисунок 3

Выполнение студентом лабораторной работы заключается в пятидесятикратном нажатии кнопки «Пуск» и регистрации в таблице измерений показаний микросекундомера. Последующая обработка результатов измерений производится в соответствии с требованиями методических рекомендаций.

## 4. Изучение распределения Больцмана

Объект исследования в данной работе – облако электронов, находящихся между катодом и анодом электровакуумного диода. Эмитированные катодом электроны – идеальный газ, температура которого равняется температуре катода. В отсутствии электрического поля концентрация электронного газа постоянна. В присутствии внешнего силового поля концентрация становится различной в разных точках пространства. Обратное включение диода (на катод подается положительный потенциал, а на анод – отрицательный) приводит к тому, что концентрация электронов от катода к аноду убывает согласно распределению Больцмана.

Шаблон компьютерной симуляции наряду с управляющими элементами содержит принципиальную схему установки, Рисунок 4. Схема включает амперметр, вольтметр, измеряющий напряжение между анодом и катодом, и реостат.

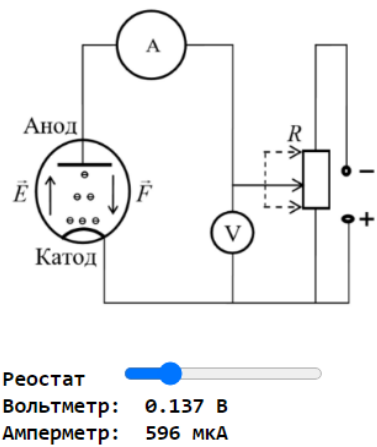

Рисунок 4

Ползунковым реостатом можно изменять показания вольтметра $U$ от 0 до 700 мВ с шагом 1 мВ.

Анодный ток $I$ при этом изменяется согласно выражению

$$I = I_{max} \exp(-qU/kT) \quad .$$

Типичная зависимость анодного тока от напряжения в полулогарифмическом масштабе приведена на Рисунке 5. Именно такую зависимость должен построить студент по результатам измерений.

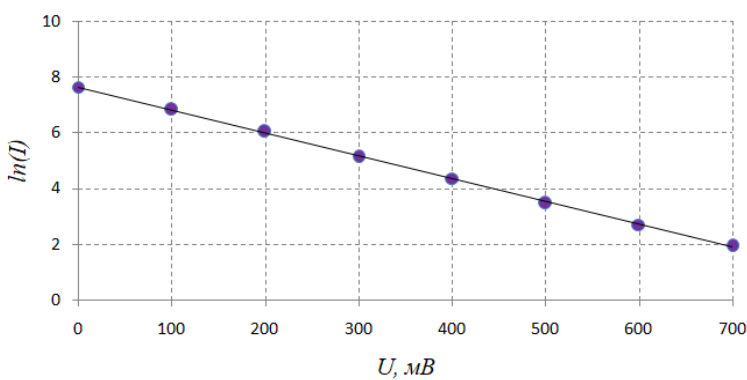

Рисунок 5

В каждой лабораторной работе индивидуальны значения силы тока в отсутствии электрического поля $I_{max}$ и температура электронного газа $T$. Диапазон рабочих значений силы тока 2100-2400 мкА, а температуры – 1100-1400 К.

Персонификация студента осуществляется по значению силы тока, это открытый параметр, температура рассчитывается в ходе выполнения работы и однозначным признаком быть не может.

Для придания естественности измерениям расчетные показания амперметра модифицируются. Значения анодного тока *I* заменяются значениями, полученными генерацией случайных чисел, равномерно распределенных в интервале *(0.8I;1.2I)*. Исключение составляет лишь используемое для персонификации значение $I(0) = I_{\max}$, оно остается неизменным.

В процессе выполнения данной лабораторной работы студент, перемещая бегунок реостата, регистрирует показания амперметра и вольтметра. Построив вольтамперную характеристику, рассчитывает температуру электронного газа.

## 5. Изучение работы источника постоянного тока

Неразветвленная электрическая цепь постоянного тока состоит из источника тока и резистора с переменным сопротивлением. Измерительные приборы, амперметр и вольтметр подключены, как показано на схеме (Рисунок 6). Зависимость падения напряжения на резисторе от силы тока $U = \varepsilon - Ir$ линейна. Сняв вольтамперную характеристику, можно определить электродвижущую силу источника, его внутреннее сопротивление, ток короткого замыкания $I_{кз} = \varepsilon/r$, полезную и полную мощность.

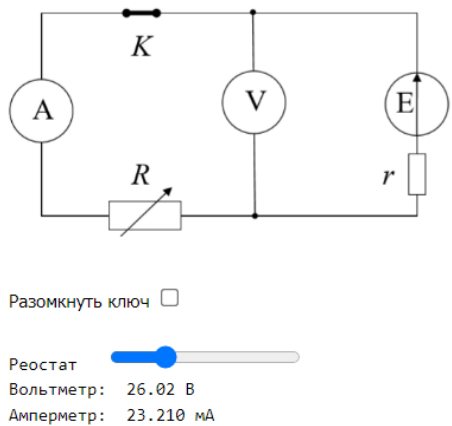

Рисунок 6

В рамках симуляции принято, что диапазон изменения ЭДС составляет 10-60 В с шагом 0.5 В, а диапазон изменения внутреннего сопротивления 400-700 Ом. Тогда диапазон тока короткого замыкания 15-150 мА.

Перемещая ползунок реостата, задаем дискретное вещественное число $n_i$, лежащее в интервале [0;1] с шагом 0.001. Тогда значения силы тока определяется выражением $I_i = I_{кз}n_i$, а напряжения $U_i = \varepsilon - rI_i$.

Итого в симуляции можно получить 1000 значений измеряемых параметров. Если номер *i>900*, то регистрируется перегрузка по току и вместо показаний амперметр выводит сообщение об этом. Если значение *i<100*, то значения берутся в 100 точке, что соответствует остаточному сопротивлению реостата. При значении *i>300* производится модификация показаний вольтметра в диапазоне ±1% от расчетного значения посредством генератора случайных чисел.

Персонифицированными и передаваемыми в шаблон симуляции для конкретного студента являются величина ЭДС (открытый параметр) и внутреннее сопротивление источника (закрытый параметр). Идентификация студента осуществляется по показаниям вольтметра при разомкнутом ключе, соответствующим значению ЭДС.

## 6. Определение удельного заряда электрона

Центральный элемент лабораторной установки для измерения удельного заряда электрона – магнетрон, электровакуумный триод, помещенный в соленоид. Зависимость анодного тока от тока соленоида позволяет рассчитать удельный заряд. Шаблон компьютерной симуляции включает принципиальную схему установки и управляющие элементы, Рисунок 7. В их число входят два амперметра, один в цепи соленоида, другой в анодной цепи, вольтметр, измеряющий напряжение между анодом и катодом, и два реостата.

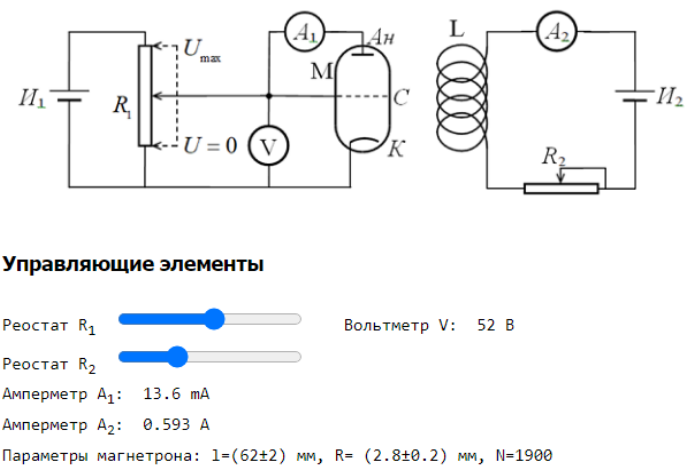

Рисунок 7

С помощью реостата $R_1$ выставляется напряжение анод-катод. Реостат $R_2$ регулирует ток в цепи соленоида от 0 до 2 А с шагом 1 мА. Анодный ток измеряется амперметром $A_1$, а ток в цепи соленоида амперметром $A_2$.

Анализ движения электронов в магнетроне приводит к следующему выводу. Идеальная зависимость анодного тока от тока в цепи соленоида $I_a(I_s)$ имеет вид ступеньки. Ток соленоида, при котором анодный ток прекращается, называется отсекающим. В условиях реального эксперимента ступенька размывается, прежде всего, по причине неоднородности магнитного поля вдоль оси соленоида конечной длины. На реальных кривых ток отсечки определяется из условия $I_a(I_{отс}) = I_0/2$.

Построим соответствие между напряжением анод-катод $U$ и током отсечки $I_{отс}$, а также анодным ток при нулевом токе в цепи соленоиде $I_0$.

Для вычисления анодного тока воспользуемся законом трех вторых

$$I_0 = aU^{3/2},$$

где $a$ – перванс, коэффициент, зависящий от конфигурации и геометрических размеров и материала электродов.

Параметры симуляции подбираются, чтобы количественно результаты опытов соответствовали режимам работы реальной установки. Если для значений перванса определить интервал $0.05 \le a \le 0.0995$, то при напряжениях (60-80) В значение анодного тока будет лежать в интервале (23-71) мА.

Пусть число витков соленоида – $N$, длина – $l$, а радиус – $R$, тогда эти параметры и удельный заряд электрона связаны соотношением

$$\frac{e}{m} = \frac{8l^2}{(K\mu_0 NR)^2} \frac{U}{I_{отс}^2} \quad .$$

Поправочный коэффициент $K$ вводится для учета неоднородности магнитного поля. Для тока отсечки идеальной ступеньки получаем $I_{отс} = b\sqrt{U}$, где $b = \frac{2l}{K\mu_0 NR}\sqrt{\frac{2m}{e}}$.

Полагая параметры магнетрона следующими: $R = (2.8 \pm 0.2)$ мм, $l = (62 \pm 3)$ мм, $K = 0.65$, и, учтя, что $\mu_0 = 4\pi \cdot 10^{-7}$ Гн/м, а $e/m = 1.75882 \cdot 10^{11}$ Кл/кг, получаем $b = 182.82885/N$.

Индивидуальными и передаваемыми в шаблон для конкретного студента являются константы $a$ и $N$. Число витков соленоида ограничено интервалом $1900 \le N \le 4000$.

Известные константы $a$ и $N$ и выставленное напряжение катод-анод в построенной модели однозначно определяют значения $I_0$ и $I_{отс}$.

Реостат $R_2$ меняет ток соленоида от 0 до 2 А с шагом 1 мА. Для расчета величины анодного тока была подобрана функция

$$I_a(I_s) = I_0 \frac{\exp\left(-\frac{I_{отс}}{D}\right) + 1}{\exp\left(\frac{I_s - I_{отс}}{D}\right) + 1},$$

которая размывает идеальную ступеньку.

Здесь $D = cI_{отс}$ – ширина области размытия, полагалось, что $c = 0.15$. Для построенной функции справедливо $I_a(0) = I_0$, ее график приведен на Рисунке 8.

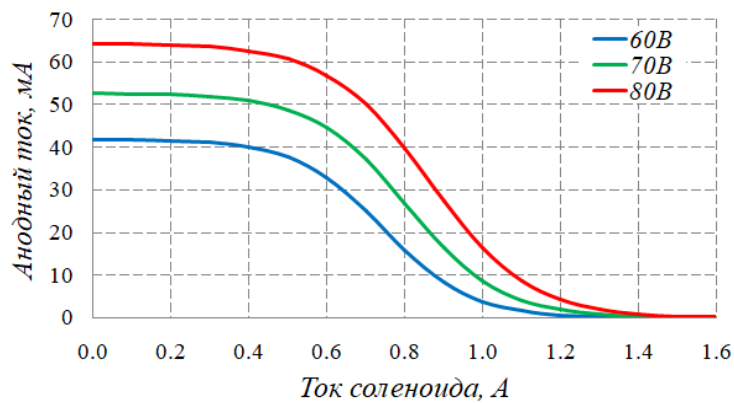

Рисунок 8

Для придания реалистичности показания амперметра $A_1$ в диапазоне (10-90)% от максимального значения зашумлялись датчиком случайных чисел.

Открытым параметром в этой работе являются число витков магнетрона, а закрытым – первеанс.

Выполнение работы заключается в том, что, выставив реостатом $R_1$ напряжение на аноде, студент, перемещая ползунок реостата $R_2$, снимает зависимость анодного тока от тока соленоида.

## 7. Вынужденные колебания в колебательном контуре

Напряжение на конденсаторе колебательного контура, содержащего последовательно включенный генератор гармонических сигналов, зависит от частоты этого сигнала. Явление резонанса можно изучать, снимая зависимость напряжения от частоты. Для измерения напряжения используется вольтметр. Принципиальная схема лабораторной установки приведена на Рисунке 9.

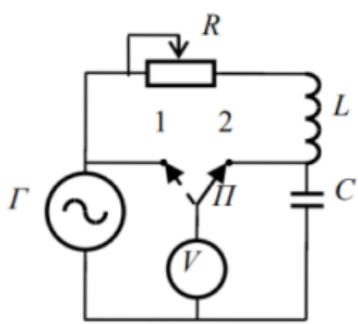

Рисунок 9

Кроме упомянутых генератора и вольтметра установка содержит реостат. Резонансные кривые снимаются при различных значениях его сопротивления.

Управляющие элементы симуляции – ползунок, изменяющий частоту выходного сигнала генератора, и ползунковый реостат, изменяющий сопротивление колебательного контура.

Индивидуальными и передаваемыми в шаблон для конкретного студента являются следующие константы: электроемкость конденсатора *C*, индуктивность соленоида *L* и три значения сопротивления, для которых снимаются резонансные кривые. Выходное напряжение генератора *E* полагается постоянным и равным один вольт.

Для того, чтобы параметры симуляции в целом соответствовали параметрам реальной установки принято, что диапазон используемых сопротивлений составляет 30-400 Ом, емкость конденсатора 20-80 нФ, индуктивность катушки 2-50 мГн. Тогда резонансная частота контура лежит в диапазоне 2920-25200 Гц.

Частоту гармонического сигнала генератора можно изменять в интервале 10-100000 Гц с шагом 10 Гц. Сопротивление ползункового реостата в колебательном контуре изменяется дискретно с шагом 1 Ом в диапазоне 10-500 Ом.

Установив заданное значение сопротивления, студент, изменяя частоту генератора ν, снимает амплитудно-частотную характеристику $U(\nu)$. Регистрируемые показания вольтметра *U* рассчитываются согласно известным формулам

$$U = \frac{\nu_0^2 E}{\sqrt{(\nu^2 - \nu_0^2)^2 + \beta^2 \nu^2 / \pi^2}} \quad , \quad \beta = \frac{R}{2L} \quad , \quad \nu_0 = \frac{1}{2\pi\sqrt{LC}} \quad .$$

Построив по результатам измерений резонансную кривую, студент определяет частоту резонанса и добротность.

8. **Заключение**

Апробация описанных работ прошла в процессе изучения физики студентами инженерных специальностей технического университета. Вынужденный переход к дистанционному обучению в условиях пандемии обусловил поиски новых форм взаимодействия преподаватель-студент. Студенты выполняли лабораторную работу, используя любое доступное им устройство. Это мог быть и планшет, и смартфон. Индивидуализация параметров виртуальной установки благотворно сказывалась на самостоятельности работы. Редкие попытки использования чужих результатов вскрывались немедленно.

Полученный в процессе работы опыт будет полезен при расширении списка симуляторов лабораторных работ.

# ЛИТЕРАТУРА

1. Лабораторные работы в домашних условиях // Кафедра общей физики /МФТИ [Электронный ресурс]. – URL https://mipt.ru/education/chair/physics/news/laboratornye_raboty_v_domashnikh_usloviyakh (дата обращения: 07.06.021).
2. Интернет-лаборатория с дистанционным управлением [Электронный ресурс]. – URL: http://remote-lab.fyzika.net/experiment/02/experiment-2.php?lng=en (дата обращения: 07.06.021).
3. Интерактивное моделирование для естественных наук и математики [Электронный ресурс]. – URL: https://phet.colorado.edu/ (дата обращения: 07.06.021).
4. Интерактивное моделирование физических явлений в среде Phet : метод. указания и сб. заданий / Новосиб. гос. техн. ун-т ; сост.: М. П. Сарина, А. В. Топовский. - Новосибирск : НГТУ, 2014. - 72 с.
5. Виртуальные лабораторные работы // Кафедра общей физики РЭФ /НГТУ [Электронный ресурс]. – URL: https://ciu.nstu.ru/kaf/of/virutalne_laboratorne_rabot (дата обращения: 07.06.021).
6. Виртуальные лабораторные работы по физике // Московский государственный образовательный комплекс (МГОК) [Электронный ресурс]. – URL: http://mediadidaktika.ru (дата обращения: 07.06.021).
7. Компьютерные симуляции лабораторных работ // Кафедра прикладной и теоретической физики / НГТУ [Электронный ресурс]. – URL: http://pitf.ftf.nstu.ru/files/zaikin/LabComSim.html (дата обращения: 07.06.021).
8. Лабораторный практикум по физике// Кафедра прикладной и теоретической физики /НГТУ [Электронный ресурс]. – URL: http://pitf.ftf.nstu.ru/resources/labs/ (дата обращения: 07.06.021).